# Stochastic Modelling and Dynamic Analysis of Cardiovascular System with Rotary Left Ventricular Assist Devices

Jeongeun Son[1], Dongping Du[2], Yuncheng Du[1*]

[1] Department of Chemical & Biomolecular Engineering, Clarkson University, Potsdam NY 13676, USA
[2] Department of Industrial, Manufacturing & Systems Engineering, Texas Tech University, Lubbock, TX 79409 USA

*Correspondence should be addressed to Yuncheng Du (email: ydu@clarkson.edu)

## Abstract

The left ventricular assist device (LVAD) has been used for end-stage heart failure patients as a therapeutic option. The aortic valve plays a critical role in heart failure and its treatment with LVAD. The cardiovascular-LVAD model is often used to investigate the physiological demands required by patients and predict the hemodynamic of the native heart supported with a LVAD. As a *"bridge to recovery"* treatment, it is important to maintain appropriate and active dynamics of the aortic valve and the cardiac output of the native heart, which requires that the LVAD pump must be adjusted so that a proper balance between the blood contributed through the aortic valve and the pump is maintained. In this paper, we investigate how the pump power of the LVAD pump can affect the dynamic behaviors of the aortic valve for different levels of activity and different severities of heart failure. Our objective is to identify a critical value of the pump power (i.e., *breakpoint*) to ensure that the LVAD pump does not take over the pumping function in the cardiovascular-pump system and share the ejected blood with left ventricle to help the heart to recover. In addition, hemodynamic often involves variability due to patients' heterogeneity and the stochastic nature of cardiovascular system. The variability poses significant challenges to understand dynamic behaviors of the aortic valve and cardiac output. A generalized polynomial chaos (gPC) expansion is used in this work to develop a stochastic cardiovascular-pump model for efficient uncertainty propagation, from which it is possible to rapidly calculate the variance in the aortic valve opening duration and the cardiac output in the presence of variability. The simulation results show that the gPC-based cardiovascular-pump model is a reliable platform that can provide useful information to understand the effect of LVAD pump on the hemodynamic of the heart.

## 1. Introduction

Cardiovascular disease is one of the major causes of death in the United States. Approximately 5.7 million adults in U. S. suffer from heart failure (HF). HF occurs when the heart fails to maintain an appropriate circulation to support the physiological demands of patient's body [1]. Heart transplantation is the well-recognized treatment for end-stage HF. However, only a few patients are eligible for transplantation, due to the limited organ donors as well as the physical limitation such as age, health condition, or other health issues (i.e., impaired renal function, other co-morbidities or a high pulmonary vascular resistance) [2]. To overcome this limitation, an alternative treatment is to implant a ventricular assist device (VAD) to help unload the ventricles. VADs are mechanical pumps, which are designed to assist either the right ventricle or the left ventricle, or both ventricles in some cases, to eject the blood into the arterial system and further into peripheral and end-organ [2].

The left ventricular assist device (LVAD) is the most commonly used device for HF patients, since the right side of the heart can often make use of the heavily increased blood flow from the LVAD. The LVAD can partially replace the mechanical work of the failing left ventricle to maintain a desired blood flow between the left ventricle and the aorta. For example, it has been used to support an ailing heart as a "*bridge to transplant*" until a suitable donor heart is available. In addition, it is considered as a "*destination therapy*" for HF patients who are not eligible for heart transplantation. Recently, LVADs have been proposed as a "*bridge to recovery*" therapeutic option to help patients recover the normal heart function [3, 4]. It was previously reported that the native heart function of patients can be improved with the support of LVADs. The reverse of HF can possibly allow patients to return to their normal life without the LVADs, and potentially improve the quality of life for HF patients [5, 6, 7].

LVADs can be generally divided into two generations by the pump types, i.e., pulsatile pumps and rotary pumps. Pulsatile pumps generate pulsatile blood flow close to the native heart in a *beat-like* fashion, whereas rotary pumps generate a continuous blood flow [8]. The rotary pump-based LVAD has more advantages over the pulsatile pumps in terms of size, efficiency, durability, noise, and weight [9]. However, an important issue with a rotary LVAD is the optimal control of pump speed, which limits its extensive use. When the pump is operated at a lower rotational speed, it can induce regurgitation (i.e., backflow) from the aorta to the left ventricle [10]. In contrast, ventricle suction can happen when the pump speed is too high, which can lead to ventricular collapse, because the pump draws more blood from the left ventricle than available. In addition, the tuning of the pump speed can inevitably affect the function of the aortic valve. For example, inappropriate selection of pump speed may lead to permanent closure of the aortic valve, which is detrimental to the cardiac recovery.

The LVAD pump is normally set at a constant speed by physicians during the implantation surgery and cannot be adjusted freely. However, the heart function changes over time and patients may have a time varying activity such as sleep, rest, and wild exercise. For both cases, the LVADs should be able to adjust the pump speed to meet different physiologic demands without inducing ventricular suction and regurgitation. The control of LVAD is a difficult problem to formulate, since physiological variables of patients with the implanted LVADs have not been well studied and the effect of changes in control variables on the cardiovascular-pump system is not well understood either. Another challenge associated with LVAD control is to assess the aortic valve dynamics, while adjusting control variable such as pump speed or pump power. To maintain a normal operation of the aortic valve for the "*bridge to recovery*", it is important to balance the amount of blood ejected through the pump and the aortic valve in order to avoid the LVAD dominates the blood circulation and takes over the heart function.

The aortic valve opens or closes periodically to allow the blood flows from the left ventricle to the aorta in each cardiac cycle. The aortic valve opens when the left ventricular pressure (*LVP*) is larger than the aortic pressure (*AoP*). As the blood flows out of left ventricle, the *LVP* decreases and the aortic valve closes. However, when the pump speed is too high, the LVAD will provide the majority of left ventricle unloading, thus the left ventricle cannot generate a sufficient pressure to open the aortic valve [11]. Consequently, the LVAD will bypass the left ventricle and the aortic valve will be closed permanently. This can significantly change the circulation physiology and introduce complications such as thrombosis and commissural fusion [12, 13]. The complications are fatal to HF patients, especially when the LVAD is used as a "*bridge to recovery*". It is important to ensure that the aortic valve can remain active, when the pump speed of LVAD is adjusted.

Mathematical models of the human cardiovascular circulatory system have been developed to understand cardiac hemodynamic. Most of these models integrate left and right ventricles and atria with the systemic and pulmonary arterial and venous system to prediction heart functions [14, 15, 16, 17]. Although these models have the potential to predict dynamic changes of the physiological states, such as the left ventricular pressure, their clinical applications are still restricted, since these models fail to consider the interacting subsystems and networks in the cardiovascular system. In addition, the cardiac function varies between different individuals and is different for the same patient over time. These variations, i.e., the *inter-* and/or *intra-patient* variability, defined as uncertainty, pose a major challenge on the development of an accurate model for the cardiovascular-pump system [18].

To improve the reliability and credibility of models, it is important to consider the uncertainty in cardiovascular circulatory system. It should be noted that sampling-based techniques such as Monte Carlo (MC) simulations are the most popular method for uncertainty analysis [19]. However, MC-based methods can be computationally demanding for modelling and control of the cardiovascular circulatory system with implanted LVAD, since a larger number of simulations are often required to obtain accurate results [20]. Recently, a generalized Polynomial Chaos (gPC) expansion-based uncertainty quantification and propagation has been studied in different modelling, optimization, and control problems [21, 22, 23, 24, 25]. The gPC-based method can propagate a probabilistic uncertainty onto model predictions in a real-time manner, from which the uncertainty in model predictions can be easily estimated from gPC coefficients [22]. Due to the computational efficiency of the gPC, it is chosen for the uncertainty analysis in the cardiovascular-pump system in this work.

Following the discussion above, a stochastic model of the human cardiovascular-pump system is developed in this work, using the gPC theory. In the presence of uncertainty, the dynamic behaviours of the aortic valve will be investigated for different electric powers. Note that the electric powers can be adjusted to vary the pump speed to meet various physiologic demands. The main contribution of this work is to efficiently quantify the uncertainty in cardiac outputs and dynamic behaviours of the aortic valve in each cardiac cycle of the cardiovascular-pump system. The uncertainty represents a time-varying physiologic change of patients in this work. Specifically, the aortic valve opening duration will be studied for different levels of physical activity and for different severities of HF. In addition, a probability description of the cardiac output such as the mean and the variance can be rapidly calculated using the gPC model, while taking uncertainty into account. It is important to note that the cardiac output can be determined by heart rate (HR) and stroke volume [10]. The stroke volume depends on the preload, contractility, and afterload of the heart. It is recognized that the rotary pump has poor sensitivity to the preload of ventricle that is related to ventricular filling with venous blood and high sensitivity to the afterload of ventricle that is the resistance to systolic ejection of blood [26, 27]. Therefore, we mainly focus on the analysis of the aortic valve dynamics and cardiac output in the presence of uncertainty in the afterload, i.e., systemic vascular resistance (SVR). The stochastic model can take into account the uncertainty among patients and individual patients' physiological activities, which lays a firm foundation for control design of LVADs.

This paper is organized as follows. Section 2 presents a deterministic cardiovascular-LVAD model and the theoretical background of the gPC theory. The results of computer simulations for the stochastic cardiovascular-pump model are presented in Section 3, which is followed by conclusions in Section 4.

## 2. Mathematical Background and Model

2.1 Cardiovascular-LVAD Model

The deterministic model of the cardiovascular-LVAD system in this work was experimentally validated by comparing the hemodynamic waveforms with data of patients [16, 28, 29]. It is assumed that the patient has a healthy and normal right ventricle and pulmonary system for simplicity. Thus, its effect on the LVAD is negligible [16, 13]. Figure 1 shows a schematic of the circuit model of the cardiovascular-LVAD system. The model parameters and their corresponding values are listed in Table 1.

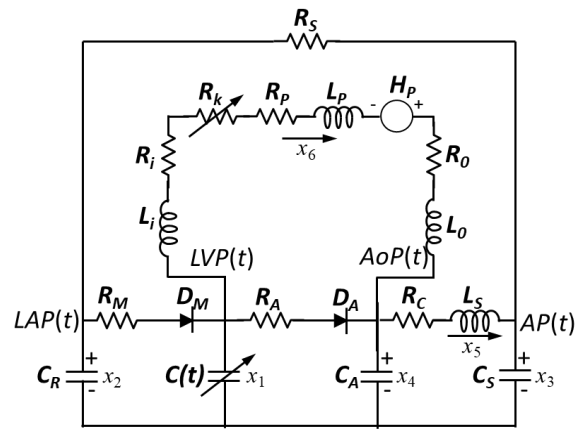

Figure 1: Cardiovascular-LVAD circuit model

Table 1: Model parameters for the cardiovascular-LVAD system

| Circuit parameters | Value | Physiological Meaning |
|---|---|---|
| Resistance (mmHg·s/ml) | | |
| $R_s$ | 1.0000 | Systemic Vascular Resistance |
| $R_M$ | 0.0050 | Mitral Valve Resistance |
| $R_A$ | 0.0010 | Aortic Valve Resistance |
| $R_c$ | 0.0398 | Characteristic Resistance |
| $R_i$ | 0.0677 | Inlet Pump Resistance |
| $R_o$ | 0.0677 | Outlet Pump Resistance |
| $R_p$ | 0.17070 | Pump Resistance |
| $R_k$ | See Equation 1 | Suction Resistance |
| | | $\alpha$ = -3.5 s/ml and $\bar{x}_1$ = 1 mmHg |
| Compliance (ml/mmHg) | | |
| $C(t)$ | Time-varying | Left Ventricular Compliance |
| $C_R$ | 4.4000 | Left Atrial Compliance |
| $C_s$ | 1.3300 | Systemic Compliance |
| $C_A$ | 0.0800 | Aortic Compliance |
| Inertances (mmHg·s$^2$/ml) | | |
| $L_s$ | 0.0005 | Inertance of Blood in Aorta |
| $L_i$ | 0.0127 | Inlet Inertance |
| $L_o$ | 0.0127 | Outlet Inertance |
| $L_p$ | 0.02177 | Pump Inertance |
| Valves (no units) | | |
| $D_M$ | / | Mitral Valve |
| $D_A$ | / | Aortic Valve |

As seen in Figure 1, the compliance $C_R$ represents the preload and pulmonary circulations and the resistors $R_M$ and $R_A$ are used to define the resistance related to mitral valve and aortic valve, respectively. The $D_M$ and $D_A$ are two ideal diodes to describe different phases in a cardiac cycle, i.e., (i) Isovolumic Relaxation, (ii) Filling, (iii) Isovolumic Contraction, and (iv) Ejection. For example, the mitral and aortic valves of a healthy heart are closed during the isovolumetric relaxation or isovolumetric contraction. The $C_A$ is the aortic compliance, and parameters $R_c$, $L_s$, $C_s$, and $R_s$ are used to describe the afterload in a four-element Windkessel model. It is important to note that the $R_s$, i.e., systemic vascular resistance (SVR), varies with respect to the level of physical activity of the patient. For example, when a person starts to do mild exercise from resting, $R_s$ will decrease and the cardiac output will increase. Similarly, the value of $R_M$, i.e., mitral valve resistance in Figure 1, will change with respect to the preload condition of the heart. The inlet and outlet resistances and inertances of the pump cannulae are described by $R_i$, $R_o$, $L_i$, and $L_o$, respectively. In addition, the pump resistance is defined as $R_p$, and the suction resistance $R_k$, illustrating the phenomenon of suction, can be defined with two parameters as below:

$$R_k = \begin{cases} 0, & if\ x_1(t) > \bar{x}_1 \\ \alpha(x_1(t) - \bar{x}_1), & if\ x_1(t) \leq \bar{x}_1 \end{cases} \quad (1)$$

where $\alpha$ indicates a cannula dependent weighting parameter, and $\bar{x}_1$ is predetermined threshold pressure of the left ventricle.

For the cardiovascular-pump system, the left ventricle compliance $C(t)$ is a time varying parameter, which is used to describe the contractibility of the left ventricle. In addition, the inverse of $C(t)$ is the elastic function of the left ventricle, i.e., $E(t)$, which can be described by the pressure and volume of the left ventricle as:

$$E(t) = \frac{1}{C(t)} = \frac{LVP(t)}{LVV(t) - V_0} \quad (2)$$

where $LVP(t)$ and $LVV(t)$ represent the pressure and volume of left ventricle, respectively. The $V_0$ is the theoretical volume at zero pressure defined as a reference volume of ventricle. The mathematical expression of the elastance function $E(t)$ used in this work is described by a "double hill" function $E_n(t_n)$ as below [16, 30]:

$$E(t) = (E_{max} - E_{min})E_n(t_n) + E_{min} \quad (3)$$

$$E_n(t_n) = 1.55 \left[ \frac{\left(\frac{t_n}{0.7}\right)^{1.9}}{1 + \left(\frac{t_n}{0.7}\right)^{1.9}} \right] \left[ \frac{1}{1 + \left(\frac{t_n}{1.17}\right)^{21.9}} \right] \quad (4)$$

It is important to note that $t_n$ is defined as: $t_n = t/(0.2+0.15t_c)$ in Equation 4, and $t_c$ represents the cardiac cycle, which is a sequence of events that occurs in each heart beat and is related to the heart rate (HR). Additionally, the elastance function $E(t)$ has different values corresponding to different heart conditions or the severities of heart failure [13]. For example, the maximum value of the elastance function, $E_{max}$, can be set to 2 mmHg/ml, which means that the elastance and the compliance function can describe the dynamic behaviour of a healthy heart. In contrast, if the value of $E_{max}$ is less than 2, it represents an unhealthy heart. Different values of $E_{max}$ can be used, depending on the severity of the heart failure (HF). For clarity, Figure 2 shows the simulation results of the elastance function $E(t)$ for three values of $E_{max}$, i.e., a healthy heart with a value of 2 mmHg/ml, a mild heart failure (HF) with a value of 1 mmHg/ml, and a severe failing heart with a value of 0.5 mmHg/ml.

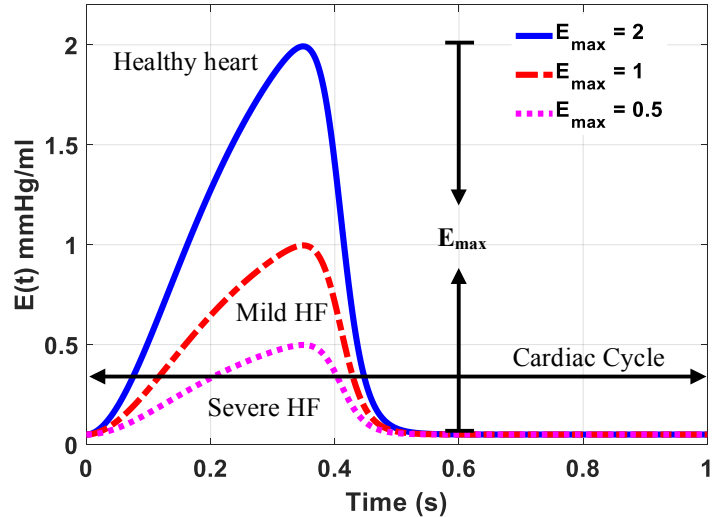

Figure 2: Elastance function of the left ventricle with respect to different severities of heart condition (cardiac cycle = 60/HR, where HR is the heart rate)

As mentioned above, the right ventricle and pulmonary system of the patient with LVAD are assumed to be healthy in this work. Thus, the blood returning from pulmonary system enters the left atrium and subsequently flows back into the left ventricle when the mitral valve open. In addition, it is assumed that some portion of the blood in the left ventricle is ejected by the LVAD and the rest of the blood is pumped by the native heart through aortic valve to the aorta and arterial system. In order to describe the cardiovascular-pump system, six state space variables, as shown in the circuit model in Figure 1, are defined in Table 2.

Table 2: State variables used in the cardiovascular-LVAD model

| Circuit Variables | Physiological variables | Physiological Meaning | Units |
|---|---|---|---|
| $x_1(t)$ | $LVP(t)$ | Left Ventricular Pressure | mmHg |
| $x_2(t)$ | $LAP(t)$ | Left Atrial Pressure | mmHg |
| $x_3(t)$ | $AP(t)$ | Arterial Pressure | mmHg |
| $x_4(t)$ | $AoP(t)$ | Aortic Pressure | mmHg |
| $x_5(t)$ | $Q_T(t)$ | Total Flow Rate | ml/s |
| $x_6(t)$ | $Qp(t)$ | Pump Flow Rate | ml/s |

Based on the basic circuit analysis, a 6$^{th}$ order state space cardiovascular-pump model can be formulated to describe the hemodynamic of a HF patient implanted with a LVAD as:

$$\dot{x} = A(t)x + P(t)p(x) + bu(t) \tag{5}$$

where $x$ is a vector of 6 state variables given in Table 2, $A(t)$ and $P(t)$ are $6 \times 6$ and $6 \times 2$ time-varying matrices, respectively, which can be defined as:

$$A(t) = \begin{bmatrix} \dfrac{-\dot{C}(t)}{C(t)} & 0 & 0 & 0 & 0 & \dfrac{-1}{C(t)} \\ 0 & \dfrac{-1}{R_S C_R} & \dfrac{1}{R_S C_R} & 0 & 0 & 0 \\ 0 & \dfrac{1}{R_S C_S} & \dfrac{-1}{R_S C_S} & 0 & \dfrac{1}{C_S} & 0 \\ 0 & 0 & 0 & 0 & \dfrac{-1}{C_A} & \dfrac{1}{C_A} \\ 0 & 0 & \dfrac{-1}{L_S} & \dfrac{1}{L_S} & \dfrac{-R_C}{L_S} & 0 \\ \dfrac{1}{L^*} & 0 & 0 & \dfrac{-1}{L^*} & 0 & \dfrac{-R^*}{L^*} \end{bmatrix} \tag{6}$$

$$P(t) = \begin{bmatrix} \dfrac{1}{C(t)} & \dfrac{-1}{C_R} & 0 & 0 & 0 & 0 \\ \dfrac{-1}{C(t)} & 0 & 0 & \dfrac{1}{C_A} & 0 & 0 \end{bmatrix}^T \tag{7}$$

It is important to note that a $2 \times 1$ vector $p(x)$ in Equation 5 is used to mimic the nonlinear behaviours of mitral and aortic valves, represented by two diodes in the circuit model in Figure 1. The details about this vector are given in Equations 8 and 9. In addition, $b$ in the last term in Equation 5 is expressed by a $6 \times 1$ vector as given in Equation 10. The control variable in Equation 5 is $u(t) = P_E(t)$, where $P_E(t)$ is the electric pump motor power, which can be regulated to adjust the pump speed of LVAD to meet the physiologic demands (e.g., blood) with respect to different levels of activity. The autonomous tuning of the pump speed is not discussed in this work, since our objective is to evaluate the distribution of the blood ejected through the LVAD and the aortic valve as a function of the electric power in the presence of uncertainty, e.g., variations in the systemic vascular resistance (SVR).

$$p(x) = \begin{bmatrix} \dfrac{1}{R_M} r(x_2 - x_1) \\ \dfrac{1}{R_A} r(x_1 - x_4) \end{bmatrix} \tag{8}$$

$$r(\zeta) = \begin{cases} \xi, & if \ \zeta \geq 0 \\ 0, & if \ \zeta < 0 \end{cases} \tag{9}$$

$$b(x) = \begin{bmatrix} 0 & 0 & 0 & 0 & 0 & \dfrac{\delta}{L^* x_6} \end{bmatrix}^T \tag{10}$$

In addition, the pressure gain (or pump head) $H_P$ (see Figure 1) can be defined as a function of the pressure difference across the pump and pump flow $x_6$ as shown in Equation 11:

$$x_1 - x_4 = R^* x_6 + L^* \frac{dx_6}{dt} - H_P \tag{11}$$

where the pump head $H_P$ can be approximated using the pump speed $\omega$ as follows [13]:

$$H_P = \beta \omega^2 \tag{12}$$

Note that the coefficient $\beta$ in Equation 12 is set to 9.9025×10$^{-7}$ mmHg/(rpm)$^2$ [13]. In Equation 11, the total resistance $R^*$ and inductance $L^*$ are two parameters related to the LVAD, which can be given by the following equations as:

$$R^* = R_i + R_o + R_p + R_k \tag{13}$$

$$L^* = L_i + L_o + L_p \tag{14}$$

The pressure gain across the pump $H_P$ has the direct relation to the electric power delivered to the pump motor $P_E$. Defining the pump efficiency as $\eta$, the electric power $P_E$ can be related to the hydrodynamic power $P_P$ as shown in Equation 15. Further, $P_P$ can be approximated with the density of the reference fluid $\rho$, gravitational acceleration $g$, the pump flow rate $x_6$, and the pump head $H_P$, as shown in Equation 16.

$$P_P = \eta P_E \tag{15}$$

$$P_P = \rho g H_P x_6 \tag{16}$$

Rearranging both equations above, the relation between the pressure gain $H_P$ and the electric power $P_E$ can be defined as follows [31]:

$$H_p = \frac{\delta P_E}{x_6} \tag{17}$$

where the constant $\delta$ is set to 7495 mmHg·ml/s·W, which is a function of the density of the reference fluid $\rho$, the gravity acceleration $g$, and the pump efficiency $\eta$ [26]. For simplicity, $\eta$ is set to 100% in this work.

Based on the definition of pump head $H_p$ in Equations 12 and 17, the corresponding rotational pump speed $\omega$ can be defined in terms of the pump electric power $P_E$ as:

$$\omega = \sqrt{\frac{\delta P_E}{\beta x_6}} \tag{18}$$

To study the effect of uncertainty in aortic valve dynamics and cardiac outputs, the operating power range of the LVAD pump is set to 0.12 to 1.56 *W* in this work [32]. The corresponding rotational pump speeds for different electric powers are shown in *Appendix A*.

2.2 Generalized polynomial chaos (gPC) expansion

The generalized polynomial chaos (gPC) expansion generally approximates a random variable using an arbitrary probability density function (PDF) defined by another random variable (e.g., $\xi$) with *a prior* distribution (PDF) in the Wiener-Askey framework [22]. In this work, the gPC expansion will be used to approximate the uncertainty such as the systemic vascular resistance SVR (i.e., $R_s$) to study and quantify how uncertainty can affect the model predictions, e.g., the aortic valve dynamics and cardiac outputs. Note that SVR in the model is used to describe the different levels of activity, which can vary over time within the same HF patient. Thus, it is assumed that the exact value of SVR is unknown at each time; however, the PDF of SVR over a period of time is available, which can be determined by physicians or estimated through offline estimation techniques. The rationale of choosing the SVR as the uncertain source in this work will be explained in the results session through the sensitivity analysis.

Suppose that the cardiovascular-LVAD system in Equation 5, described by a set of nonlinear ordinary differential equations (ODEs), can be simplified as follows:

$$\dot{x} = f(t, x, v, g; u) \tag{19}$$

where the vector $x$ consists of the 6 state variables defined in Table 2 with initial values $x_0$ at $t = 0$. The $v$ is a vector of deterministic model parameters in the cardiovascular-pump system, which are fixed constants. In contrast, a vector of parametric uncertainties is defined by $g$, which will be approximated with their PDFs instead of using fixed constants, such as the SVR explained above, representing the level of activity of HF patients. In addition, $u$ is the control variable, e.g., electric power $P_E(t)$, which can be adjusted to meet the physiologic demands.

To evaluate the effect of uncertainty on the model predictions $x$, each parameter $g_i$ ($i$=1,2, ..., $n_g$) in $g$ will be approximated with a gPC model as a function of a set of independent random variable $\xi = \{\xi_i\}$ as:

$$g_i = g_i(\xi_i) \tag{20}$$

where $\xi_i$ is the $i^{th}$ random variable used to approximate $g_i$ that follows *a prior* PDF defined in the Wiener-Askey framework. Since each parametric uncertainty is approximated with a gPC model, the model predictions $x$ can be also defined with random variables $\xi$, which can assess the effect of uncertainty on the model predictions of the cardiovascular-pump system. Using the orthogonal polynomial basis functions defined in the Wiener-Askey framework, the gPC approximations for both uncertainty and model predictions can be defined as follows:

$$g_i(\xi) = \sum_{k=0}^{\infty} \hat{g}_{i,k}\, \varphi_k(\xi_i) \approx \sum_{k=0}^{q} \hat{g}_{i,k} \varphi_k(\xi_i) \tag{21}$$

$$x_j(t, \xi) = \sum_{k=0}^{\infty} \hat{x}_{j,k}(t) \psi_k(\xi) \approx \sum_{k=0}^{Q} \hat{x}_{j,k}(t) \psi_k(\xi) \tag{22}$$

where $\{\hat{g}_{i,k}\}$ and $\{\hat{x}_{j,k}(t)\}$ are the gPC coefficients of the $i^{th}$ parametric uncertainty and the $j^{th}$ model prediction (state variable), and $\varphi_k(\xi_i)$ and $\psi_k(\xi)$ are multi-dimensional polynomial basis functions. When the PDF of $g_i$ is given or can be estimated so that $g_i(\xi)$ follows *a prior* distribution, the gPC coefficients $\{\hat{g}_{i,k}\}$ in Equation 21 can be subsequently determined. As compared to $\{\hat{g}_{i,k}\}$, the gPC coefficients of $\hat{x}_{j,k}(t)$ will be calculated by substituting Equations 21 and 22 into Equation 5, which is followed by using a Galerkin projection. The Galerkin projection between a state variable $x_j$ and a polynomial basis function can be defined as:

$$\langle \dot{x}_j(t, \xi), \psi_k(\xi) \rangle = \langle f(t, x_j(t, \xi), v, g(\xi); u), \psi_k(\xi) \rangle \tag{23}$$

For each state variable $x_j$, the Galerkin projection will produce a system with a set of coupled deterministic equations, in which the first equation provides the mean value of $x_j$ at each time instant, while the rest equations can be used to estimate the variation resulting from uncertainty. In addition, as seen in Equations 21 and 22, infinite number of terms is often truncated into a finite number of terms for practical application, i.e., $q$ and $Q$, respectively. The number of terms in Equation 21, i.e., $q$, can be optimally selected, so that the gPC model of uncertainty can approximate *a prior* known PDF of uncertainty. The total number of terms $Q$ in Equation 22 can be computed with a heuristic formula, which can be defined by the polynomial order $q$ and the total number of parametric uncertainty $n_g$ as:

$$Q = ((n_g + q)!/(n_g! q!)) - 1 \qquad (24)$$

In addition, the inner product between any two vectors in Equation 23 can be determined as:

$$\langle \phi(\xi), \phi'(\xi) \rangle = \int \phi(\xi)\, \phi'(\xi) W(\xi) d\xi \qquad (25)$$

where the integration on the right-hand side of Equation 25 is performed over the entire domain defined by random variables $\xi$, and $W(\xi)$ is the weighting function, i.e., the probability density function of $\xi$, which is selected according to the polynomial basis function in the Wiener-Askey framework. For example, a normally distributed random variable $\xi$ should be used, when the uncertainty follows a normal distribution [22]. Thus, Hermite polynomial basis functions are the best choice of the weighting function.

Using the gPC coefficients of state variables **x** in Equation 22, the statistical moments such as mean and variance of **x** at a given time interval $t$ can be quickly computed as follow:

$$E(x_j(t)) = E\left(\sum_{i=0}^{Q} \hat{x}_{j,i}(t)\psi_i\right) = \hat{x}_{j,i}(t)\,E(\psi_i) + \sum_{i=1}^{Q} E(\psi_i) = \hat{x}_{j,0}(t) \qquad (26)$$

$$\begin{aligned}
Var(x_j(t)) &= E\left(x(t) - E(x_j(t))^2\right) = E\left(\left(\sum_{i=0}^{Q}\hat{x}_{j,i}(t)\psi_i - \hat{x}_{j,(i=0)}(t)\right)^2\right) \\
&= E\left(\left(\sum_{i=1}^{Q}\hat{x}_{j,i}(t)\psi_i\right)^2\right) = \sum_{i=1}^{Q} \hat{x}_{j,i}(t)^2\, E(\psi_i^2)
\end{aligned} \qquad (27)$$

As seen, the first statistical moment of **x**, i.e., mean value, can be approximated with the first gPC coefficient $\hat{x}_{j,\,i=0}$, while the second statistical moment, i.e., variance of **x**, can be calculated using the higher order gPC coefficients and the mean of squared orthogonal polynomial basis function $\psi_i(\xi)$. Once again, the variance in state variables **x** originates from uncertainty **g**, as defined in Equation 20.

The gPC provides analytical formulas to calculate the statistical moments of model predictions, from which the PDF profiles can be rapidly estimated. This is the main rationale to use the gPC approximation in this work. The fast calculation of uncertainty in hemodynamic waveforms, resulting from the time-varying physiologic change such as SVR ($R_s$) in this work, can provide useful information to better understand the physiologic demands. Specifically, the PDF profiles of **x**, approximated from a gPC model, will be used to account for the effect of uncertainty in $R_s$ on the aortic valve opening duration and the cardiac outputs, which will be discussed in details in Section 3.

# 3. Results and Discussion

3.1 Cardiac Hemodynamic

The cardiovascular-pump model as explained in Section 2.1 involves a few model parameters, which can be used to represent the physiological dynamics of the heart supported with a LVAD. As previously reported, parameters $R_s$, $R_M$, and heart rate (HR) can govern the behaviour of the cardiovascular-pump system [13]. Thus, we propose to investigate their effect on the dynamics of the aortic valve and the cardiac output.

Since the gPC-based stochastic model in this work is developed based on a deterministic model in Equation 5, the first step is to validate the ability of the deterministic model in order to mimic the hemodynamic of the cardiovascular-pump system. The linear relationship between the end-systolic pressure and the left ventricle volume in the presence of perturbations in $R_s$ and $R_M$ is used. Note that the perturbations in $R_s$ and $R_M$ represent the stochastic changes during the preload and afterload of the heart. A total of 4 preload and 4 afterload changes were simulated as shown in Figure 3, for which the left ventricle parameters such as $E_{max}$, $E_{min}$, and $V_0$ are set to constant and the electric pump power is set to $P_E = 0.12\ W$.

The left ventricle pressure and the left ventricle volume graph (PV-loop) are shown in Fig. 3. The PV-loops in Figure 3 (a) shows the hemodynamic changes resulting from variations in the afterload, i.e., different values of $R_s$ (SVR) that represent the changes in the level of activity. In contrast, the PV-loop in Figure 3 (b) shows the result by altering the preload, i.e., $R_M$ representing the mitral valve resistance. For both graphs, a linear relationship between the left ventricle pressure and the left ventricle volume is observed, thus confirming the ability of the model to describe the hemodynamic of the cardiovascular-pump system. In addition, it was found that any variations in the cardiovascular-pump system can significantly affect the PV-loop such as the area that represents the stroke work.

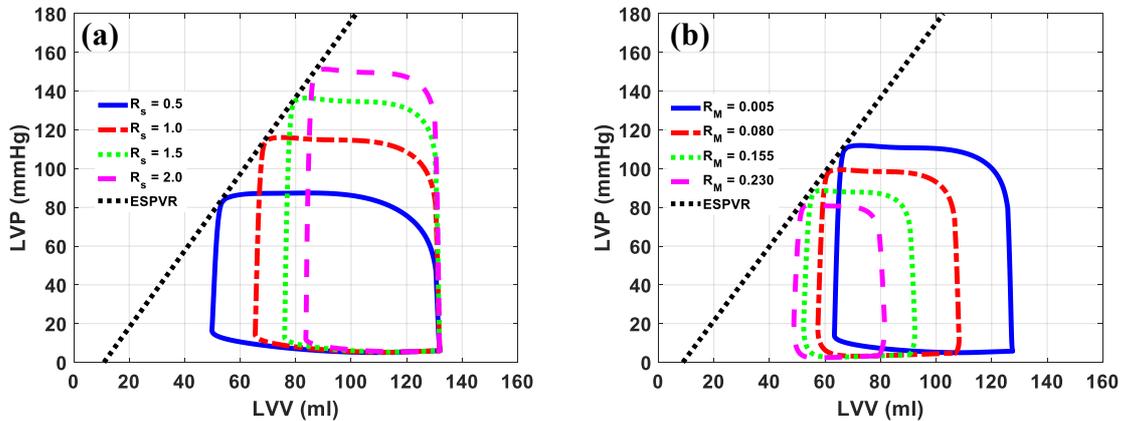

Figure 3: PV loops for different values of (a) the resistance $R_s$ (afterload) and (b) $R_M$ (preload)

Note that the dynamic behaviours of the aortic valve will be used in this work as a criterion to evaluate the effect of uncertainty on the cardiovascular-pump system. The rationale is that any changes in physiological parameters such as the $R_s$ can affect the left ventricular pressure as seen in Figure 3 (a). Constant low ventricular pressure can result in the permanent closure of the aortic valve, which is detrimental to cardiac recovery. Specifically, the aortic valve opening duration is used to quantitatively assess the effect of uncertainty on the dynamic behaviours of the aortic valve.

The calculation of the aortic valve opening duration proceeds as follows. In a cardiac cycle, the period of time, $t_{open}$, during which the left ventricular pressure ($LVP$) is larger than the aortic pressure ($AoP$) will be first determined. The aortic valve opening duration can then be obtained by calculating the ratio between $t_{open}$ and the time of a cardiac cycle $t_c$. A schematic of the calculation of the aortic valve opening duration is given in Figure 4. Let suppose there are 9 data points of both $LVP$ and $AoP$ in a cardiac cycle, and the time interval between any two data points is $t_{int}$. As seen, two data points of $LVP$ are larger than the $AoP$, indicating that $t_{open}$ is ($2\times t_{int}$). Similarly, the cardiac cycle can be calculated as $t_c = 9\times t_{int}$. Thus, the aortic valve opening duration can be determined with $2/9 \approx 0.222$, representing that the aortic valve remains open during approximately 22.2% of a cardiac cycle.

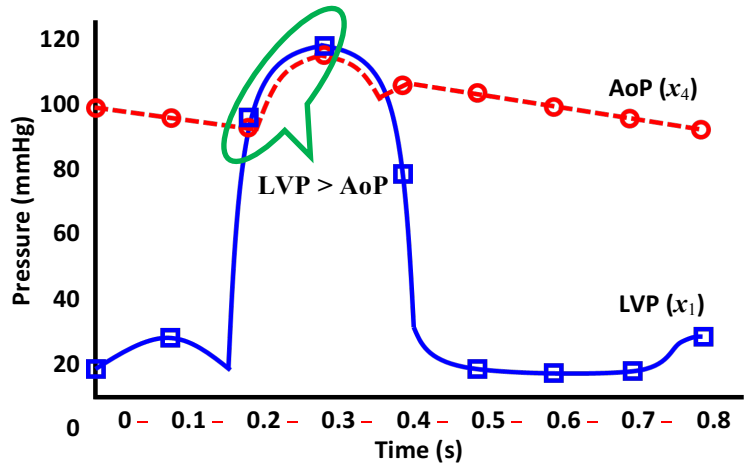

Figure 4: Schematic for the calculation of the aortic valve opening duration

The cardiac output (CO) is another key physical property, which can be calculated as a product of stroke volume and HR. The cardiac output represents a total amount of blood pumped from the ventricle in a minute. The stroke volume is the blood volume pumped from the left ventricle in a beat and is determined by the contractility, preload, and afterload. The area of the PV-loop is often calculated to represent the stroke work. As shown in Figure 3, any changes in $R_s$ and $R_M$ can affect the total amount of blood entering the ventricle, thus affecting the stroke volume and the cardiac output. Therefore, the cardiac output is used as a second criterion to evaluate the effect of uncertainty on the cardiovascular-pump system in this work.

3.2 Sensitivity Analysis

As mentioned above, $R_s$, $R_M$, and HR can affect the physiological dynamics of the heart such as the aortic valve opening and the cardiac output. However, each factor may have different effect on the hemodynamic of the failing heart. Thus, a sensitivity analysis is first performed to identify the most significant physiological factor. The effect of variations in $R_s$, $R_M$, and HR on the aortic valve and the cardiac output is first investigated, using the cardiovascular-LVAD model with a mild HF patient (i.e., $E_{max} = 1.0$) and electric pump power $P_E = 0.12 \sim 0.6\ W$. Similar results were found for other heart conditions and pump powers, but it is not shown for brevity.

It is assumed that each parameter can vary randomly between two levels, i.e., +1 and -1, which corresponds to a +10% change and a -10% change with respect to its nominal values. Note that the nominal values of each parameter are $R_s = 1.0$ mmHg·s/ml, $R_M = 0.005$ mmHg·s/ml, and HR = 75 bpm. For each parameter, let define the aortic valve opening duration (or the cardiac output) as $w_{g_i}^+$ and $w_{g_i}^-$ for two different levels, and the corresponding result with the nominal values of model parameters as $w_{g_i}^0$. Then, the effect of uncertainty in the $i^{th}$ parameter ($R_s$, $R_M$, and HR) on either the aortic valve opening duration or the cardiac output can be defined with a sensitivity index as follows:

$$\delta w_{p_i} = \frac{|w_{g_i}^+ - w_{g_i}^0|}{w_{g_i}^0} + \frac{|w_{g_i}^- - w_{g_i}^0|}{w_{g_i}^0} \tag{28}$$

where $\delta w_{p_i}$ is a sensitivity index that can be used to decide the significance of the $i^{th}$ parameter.

Based on the sensitivity index, a half-normal probability diagram is used [33] to identify the significant parameters, i.e., the factors induce significant changes in the aortic valve opening duration and the cardiac output. The half-normal probability is defined as:

$$\left[\Phi^{-1}\left(0.5 + \frac{0.5[i - 0.5]}{k}\right), \delta w_{p_i}\right] \tag{29}$$

where $i = 1, \ldots, k$ is the $i^{th}$ parameter, $\Phi^{-1}$ represents the cumulative distribution function (CDF) with respect to a standard normal distribution.

For clarity, Figure 5 shows the sensitivity analysis results for different electric pump powers $P_E$. Note that Figures 5 (a)~(c) are the results for the aortic valve opening duration, while Figure 5 (d)~(f) are the results for the cardiac output. The electric pump power in Figures 5 (a) and (d) was set to 0.12 $W$, and 0.36 $W$ was used in Figures 5 (b) and (e). For Figures 5 (c) and (f), the pump power was set to a relatively larger value of 0.6 $W$. It is worth mentioning that the pump motor power was chosen to avoid the permanent closure of the aortic valve.

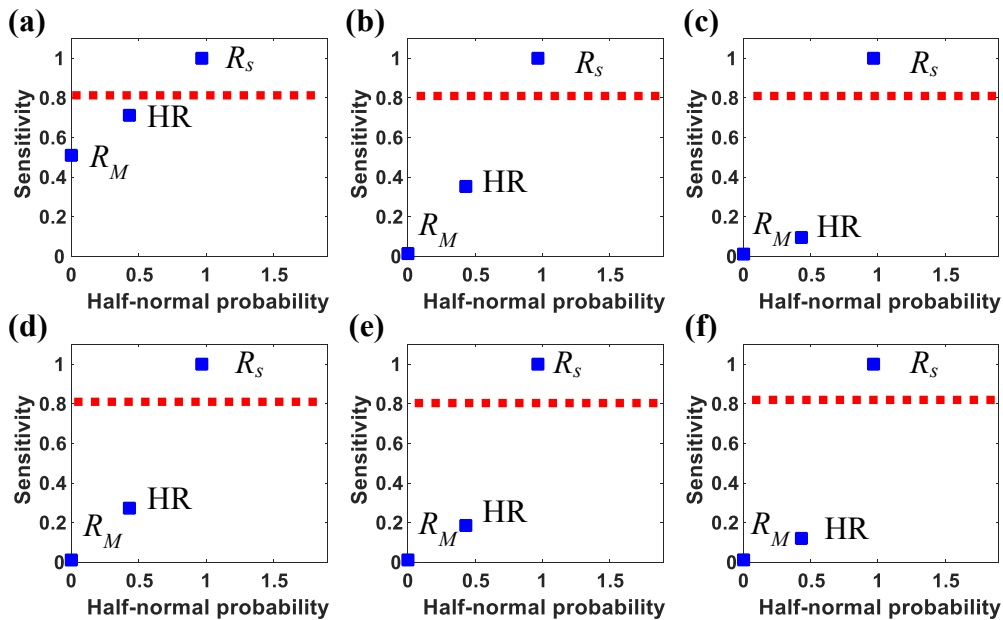

Figure 5: Half-normal probability plots for sensitivity analysis: (a) Results of aortic valve opening duration with $P_E$ = 0.12 $W$; (b) Results of aortic valve opening duration with $P_E$ = 0.36 $W$; (c) Results of aortic valve opening duration with $P_E$ = 0.6 $W$; (d) Results of cardiac output with $P_E$ = 0.12 $W$; (e) Results of cardiac output with $P_E$ = 0.36 $W$; (f) Results of cardiac output with $P_E$ = 0.6 $W$.

As seen in Figure 5, it was found that the systemic vascular resistance (SVR), i.e., $R_s$, is the most significant uncertainty for all the case studies except in Figure 5 (a). To validate the results, additional case studies were performed to evaluate the effect of variations in $R_s$ and HR on the aortic valve opening duration with respect to a relatively larger range of the pump powers, and the results were summarized in Figure 6.

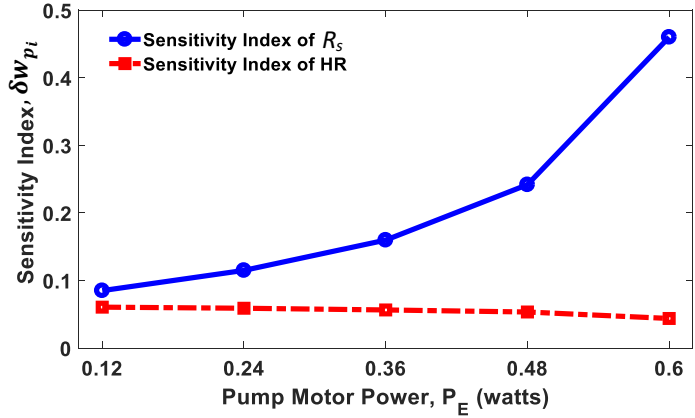

Figure 6: Sensitivity index of $R_s$ and HR corresponding to pump motor power

As seen in Figure 6, it was found that the sensitivity index of $R_s$ increases as the pump power $P_E$ increases, whereas the sensitivity index of HR decreases slowly when $P_E$ is increased. This indicates that the hemodynamic such as the aortic valve opening is more sensitive to variations in the level of activity of a HF patient, i.e., $R_s$, especially when the electric pump power is higher. Since the pump is often operated above the minimum value of the pump power, we will focus on the level of activity $R_s$ in this work for the rest of the study. It is important to note that similar results were previously reported by clinical studies, i.e., a rotary LVAD pump is more sensitive to the afterload of left ventricle [26, 27].

Based on discussion, $R_s$ is identified as the most sensitive factor in the cardiovascular-LVAD system that can affect the cardiac hemodynamic. Following the procedures as discussed in Section 2.2, a gPC model is developed to study the effect of the variations in $R_s$ on the aortic valve opening duration and the cardiac output.

3.3 Formulation of Stochastic gPC Model

It is assumed that the systemic vascular resistance (SVR), i.e., $R_s$ in this work, follows a normal distribution. Three mean values of $R_s$ are used for algorithm verification, i.e., 0.5 mmHg/ml/s, 1 mmHg/ml/s, and 2 mmHg/ml/s, which represent different levels of physical activity, i.e., very active (0.5 mmHg/ml/s), moderately active (1 mmHg/ml/s), and inactive (2 mmHg/ml/s). For example, the very active state represents that the patient is walking stairs, while the inactive state means the patient is resting or sleeping. To build a gPC model of SVR, the order of the polynomial chaos expansion is set to 1, i.e., $q = 1$ in Equation 21, since the uncertainty is normally distributed. Using Equation 24, the expansion of each state variable in Table 2 would involve 2 terms, i.e., $Q = 1$, since there is one uncertainty ($n_g = 1$) and the highest order of polynomial expansion of uncertainty is 1 ($q=1$).

To introduce perturbations in SVR, a 10% variation around each mean value is used, which can be used to determine the gPC coefficients of $R_s$ in Equation 21. The gPC coefficients of the hemodynamic variables ($x_1$-$x_6$) can be calculated by substituting the gPC models of $R_s$ and each state variable into Equation 5 and by using a Galerkin projection, from which a stochastic model can be formulated. This model can describe dynamic behaviours of the cardiovascular-pump system in the presence of uncertainty in SVR ($R_s$). For brevity, the stochastic model is presented in *Appendix B*. The simulation results of the stochastic cardiovascular-pump model are shown in Figure 7.

The first and second columns in Figure 7 show the gPC coefficients of the state variables in Equation 5, i.e., the left ventricle pressure ($x_1$), the aortic pressure ($x_4$), and the flows of pump and aorta ($x_5$ and $x_6$), respectively. For the simulations of 20 cardiac cycles, the first 3 cardiac cycles are given in the first column, whereas the second column shows the last two cardiac cycles. The third column shows the variations in these states resulting from perturbations in SVR ($R_s$). For the results shown in Figure 7, the HR was set to 75 bpm, and the mean value of $R_s$ was 1 mmHg/ml/s. The maximum elastane ($E_{max}$) was set to 1 mmHg/ml, which represents a native heart with a mild heart failure, and the pump motor power $P_E$ used in this case study was 0.12 $W$ to avoid the permanent closure of the aortic valve. It is important to note that the difference between the waveforms of the pump flow ($x_6$) and the aortic flow ($x_5$) in each cardiac cycle can be used to estimate the blood flow ejected by a native heart.

As seen in the third column of Figure 7 (c) and (f), the left ventricular pressure ($x_1$) and the aortic pressure ($x_4$) can be affected by perturbations in SVR ($R_s$). In addition, it was found that the resulting variation in the aortic pressure is more significant than the left ventricular pressure. Note that the variations in $x_1$ increase as the electric power increases, however, it is not shown for brevity. Using the gPC coefficients, it is possible to estimate the upper and lower bounds of all state variables at each time interval. For example, Figure 7 (f) shows the results of the aortic pressure, where σ is the standard deviation calculated with the higher order gPC coefficients in Equation 27, i.e., $0 < k \leq Q$. The range defined by the upper and lower bounds can quantify the dynamic values of aortic pressure within three standard deviation of the mean value of aortic pressure, which shows the 99% confidence interval of the aortic pressure at a particular time interval.

As explained in Section 3.1, the aortic valve opens when the left ventricular pressure is higher than the aortic pressure. Using the upper and lower bounds of the left ventricular pressure and the aortic pressure, the effect of perturbations in $R_s$ on the aortic valve opening duration can be quickly estimated. In addition, the mean and the variance in the pump flow can be rapidly calculated with Equations 26 and 27 as shown in Figures 7 (i) and (l), from which the variation in the cardiac output can be estimated. The calculation of the variance in the aortic valve opening duration and the cardiac output will be further discussed in the following sections.

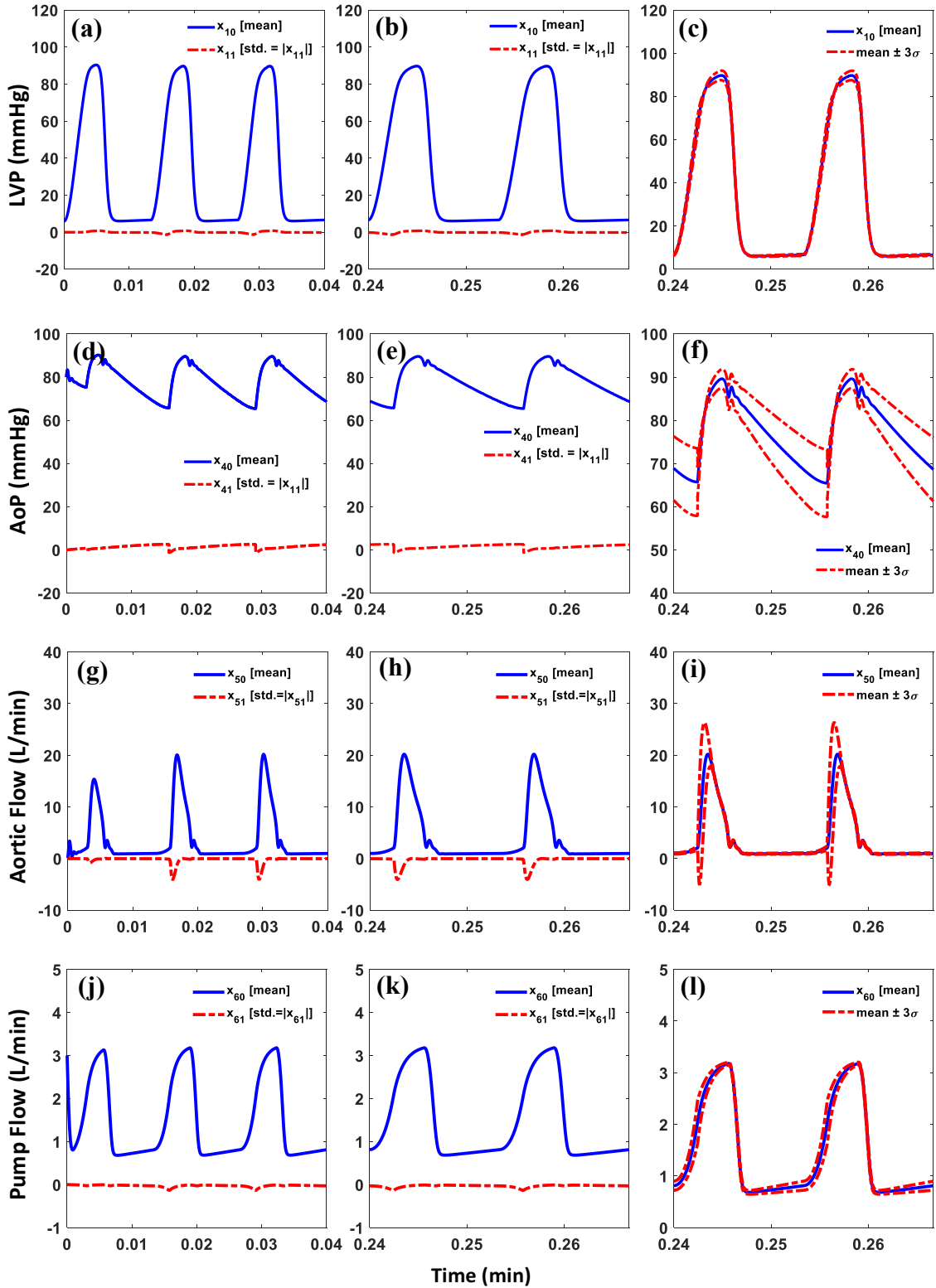

Figure 7: Hemodynamic waveforms in the presence of uncertainty in SVR. For the first and second column, the solid blue lines represent the mean value of each state variable in Equation 5, while the red dotted line represents the 2$^{nd}$ order gPC coefficients that can be used to calculate the standard deviation (σ in the last column) at each time interval. The last column shows the mean, maximum, and minimum value of each state variable, i.e., left ventricular pressure, aortic pressure, aortic flow, and the pump flow.

3.4 Dynamics of Aortic Valve Flow and Pump Flow

The effect of perturbations in the SVR ($R_s$) on the state variables has been discussed in previous section, where the electric pump power $P_E$ was fixed. In this case study, the main objective is to investigate the effect of uncertainty on the aortic valve dynamics of the heart supported by a LVAD with a time-varying pump power. For this purpose, the electric pump power $P_E$ was changed from 0.12 to 1.56 $W$ in order to better study the effect of the pump speed on the aortic valve over a wide range of operating condition. The aortic valve opening duration can be determined by examining the difference between the left ventricular pressure and the aortic pressure, as explained above [13]. When the aortic valve opens, a certain portion of the blood from the left ventricle flows through the aortic valve. As the pump motor power increases, the portion of blood flows through the aortic valve will be decreased. It was found that, when the pump power reaches a certain level, the LVAD takes over the heart function and the aortic valve can be fully by-passed, i.e., permanently closed. Figures 8 and 9 show the changes in the aortic valve flow and pump flow for different pump power values ($P_E$) in the presence of perturbations in SVR ($R_s$).

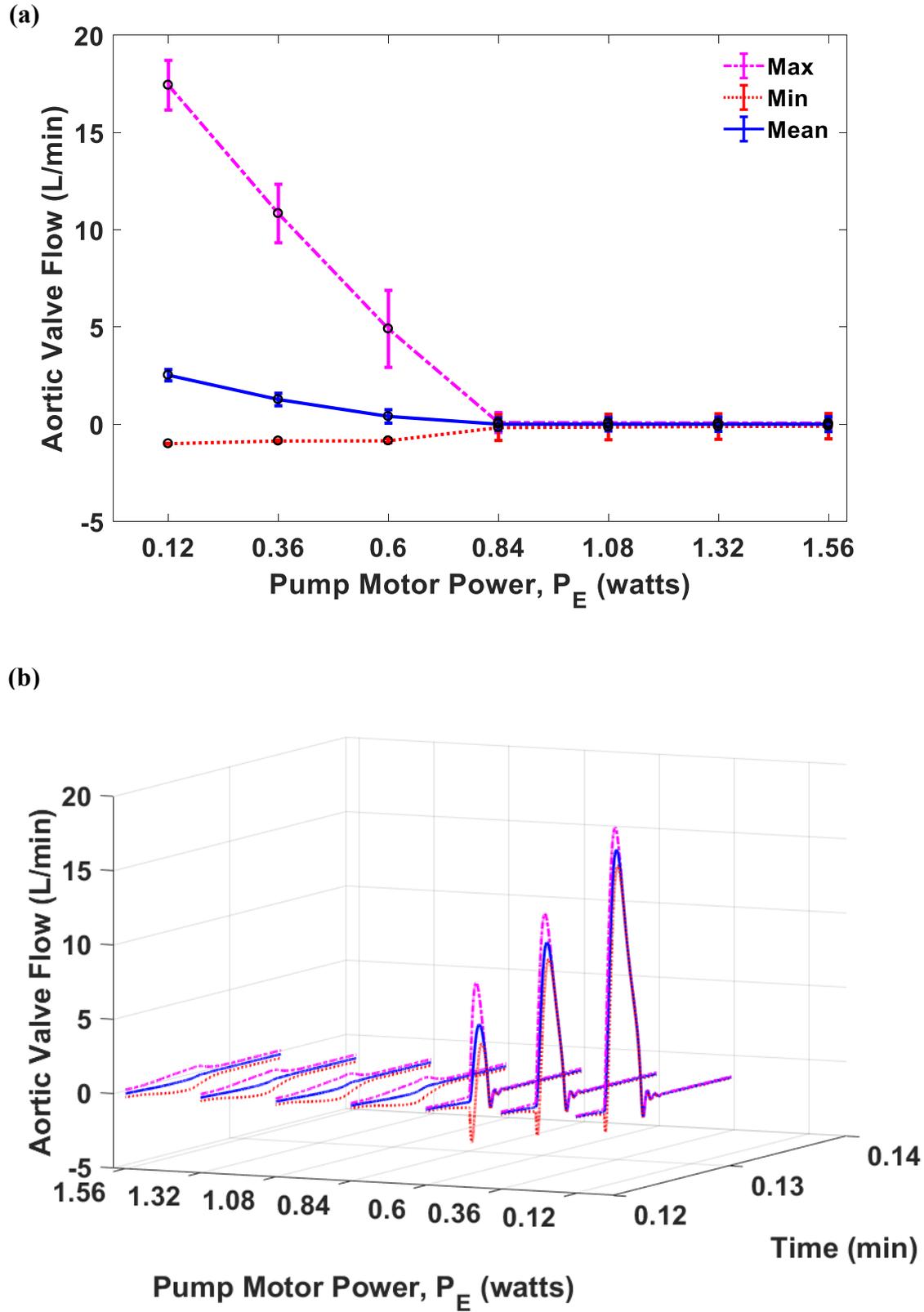

Figure 8: Hemodynamic waveforms of aortic valve flow in the presence of uncertainty in SVR ($R_s$)

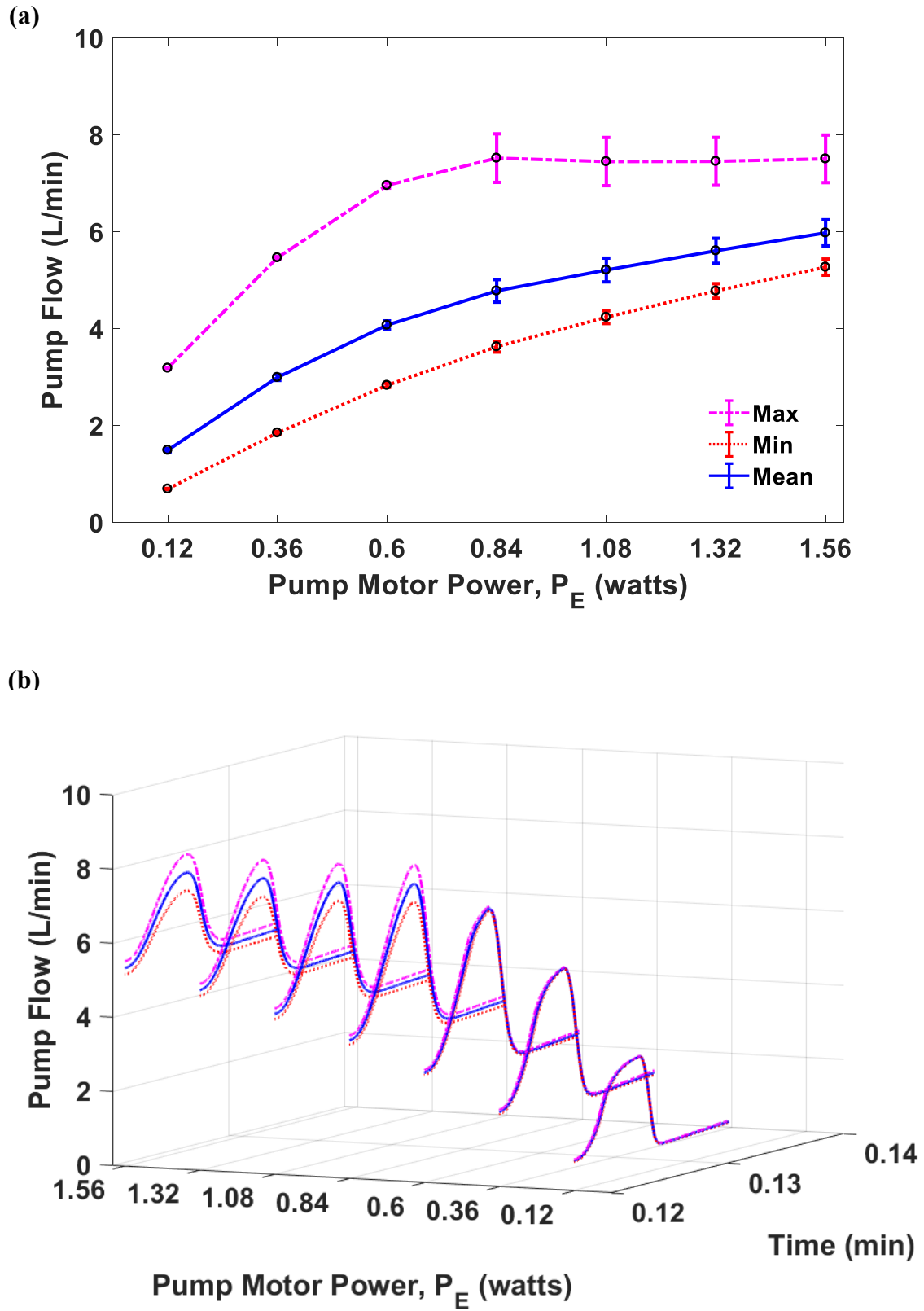

Figure 9: Hemodynamic waveforms of pump flow in the presence of uncertainty in SVR ($R_s$)

Due to the pulsatility in hemodynamic waveforms, Figures 8 (a) and 9 (a) show the maximum, minimum, and mean values of the aortic valve flow and the pump flow corresponding to different pump powers, respectively. It is worth mentioning that the maximum, minimum, and mean values of the state variables, i.e., aortic valve flow and pump flow in Figures 8 (a) and 9 (a), are calculated from the first gPC coefficient in Equation 22 of each state variable, i.e., $\hat{x}_{j,k=0}$. In contrast, the bar plot, representing the confidence interval, is calculated with other higher order gPC coefficients, i.e., $\hat{x}_{j,k}$ ($0 < k \leq Q$), using Equation 27.

Figures 8 (b) and 9 (b) show the simulation results of the hemodynamic waveforms of the aortic valve flow and the pump flow for a specific cardiac cycle, when different pump powers $P_E$ were used. The variances around maximum and minimum values in both flows through aortic valve and pump are calculated by using the high order of gPC coefficients corresponding to the maximum values, while the variances of mean values for both flows are obtained based on the upper and lower bounds quantified with standard deviation calculated by gPC coefficients. It is important to note that the pump power $P_E$ can be automatically selected according to the level of activity in order to meet different physiological demands, but this is not discussed in this work, as our objective is to evaluate the dynamic behaviors of the aortic valve over a wide range of pump power.

As seen in Figure 8 (a), it was found that the maximum and mean values of the aortic valve flow decrease as the increase of the pump motor power $P_E$. Note that the aortic valve was completely closed when the pump power $P_E$ was increased to 0.84 *W*, which can be defined as a *breakpoint*. In addition, as seen in Figure 8 (a), when the pump power $P_E$ is below the *breakpoint*, negative values of the minimum aortic valve flow was observed. This can possibly attribute to regurgitation of flow through the aortic valve, because of the aortic compliance. Physiologically, this phenomenon can be caused by the adverse pressure gradient, which can be developed as the aortic flow starts to decelerate quickly after reaching its maximum. This can eventually affect the low momentum fluid near the wall of the aorta, thus inducing reverse flow in the sinus region, which was previously reported [34]. It is important to note that the maximum and mean values of aortic valve flows remain positive. The reverse flow only exists in the minimum values of aortic valve flow, which is negligible and can be decreased as the pump power is increased. After the *breakpoint*, i.e., $P_E = 0.84$ *W*, the aortic flow is reduced to approximately 0 L/min, which means that the LVAD takes over the left ventricle function for the entire cardiac cycle and there is negligible blood that can flow through the aortic valve. Further, as can be seen in Figure 8 (b), the pulsatility of the aortic flow decreases as the pump power is increased. When the pump power $P_E$ reaches the *breakpoint*, the aortic valve flow can completely lose the pulsatility. Such information can be useful for the controller design to adjust the pump speed to meet the different physiological demands.

As shown in Figure 9 (a), the maximum, minimum, and mean values of pump flow rate increase when the pump power was increased. The maximum values of the pump flow increase until the pump power reaches the *breakpoint* and then converges to a constant after the *breakpoint* $P_E = 0.84$ *W*, while the minimum and mean values of the pump flow keep increasing as the pump power increases. In addition, it is important to note that the dashed lines (purple and red) in Figure 9 (b) represent the variations in the pump flow, resulting from the perturbations in the SVR ($R_s$). As seen, when the pump takes over the native heart, the variation in pump flows becomes larger, as compared to the cases when the aortic valve operates normally (i.e., before reaching the *breakpoint*). Further, it is worth mentioning that the variations of the aortic valve flow and pump flow shown in Figure 8 and 9 are correlated to the pump motor power. It was found that the aortic valve flow can be significantly affected by perturbations in $R_s$ before the *breakpoint*. In contrast, the variation in pump flow is larger when the LVAD pump begins to take over the left ventricular function after the *breakpoint*. This observation can be used as tuning constraints of pump power in the controller design, since larger variations in the pump flow and the aortic flow may weak the myocardium, which is detrimental to cardiac recovery and can be fatal to HF patients with LVADs. For example, constraints can be used to confine the allowable tuning range of the pump power to avoid inducing larger variations in pump flow, while taking perturbations in SVR ($R_s$) into account.

3.5 Aortic Valve Opening Duration

In Section 3.4, it was found that the aortic valve flow rate varies with respect to the electric pump power. The aortic valve can be closed when the pump power reaches a certain level (e.g., *breakpoint*), at which the LVAD pump takes over the heart function. Based on this observation, the objective in this case study is to examine the aortic valve opening duration in a cardiac cycle, while considering the perturbations in SVR ($R_s$).

As explained in Section 3.1, the aortic valve opening duration is measured by calculating the ratio between the time that the aortic valve opens during a cardiac cycle and the duration of the cardiac cycle. For clarity, the mean values of the aortic valve opening duration for a mild HF patient ($E_{max}$ = 1) and a severe HF patient ($E_{max}$ = 0.5) with different electric pump powers are shown in Figures 10 and 11, respectively. In addition, the variations around the mean values of aortic valve opening duration are summarized in *Appendix C* for brevity. In this study, HR is set to 75 bpm for both cases, and $R_s$ follows the same probabilistic description as done in Section 3.3, i.e., three different levels of physical activity of a patient.

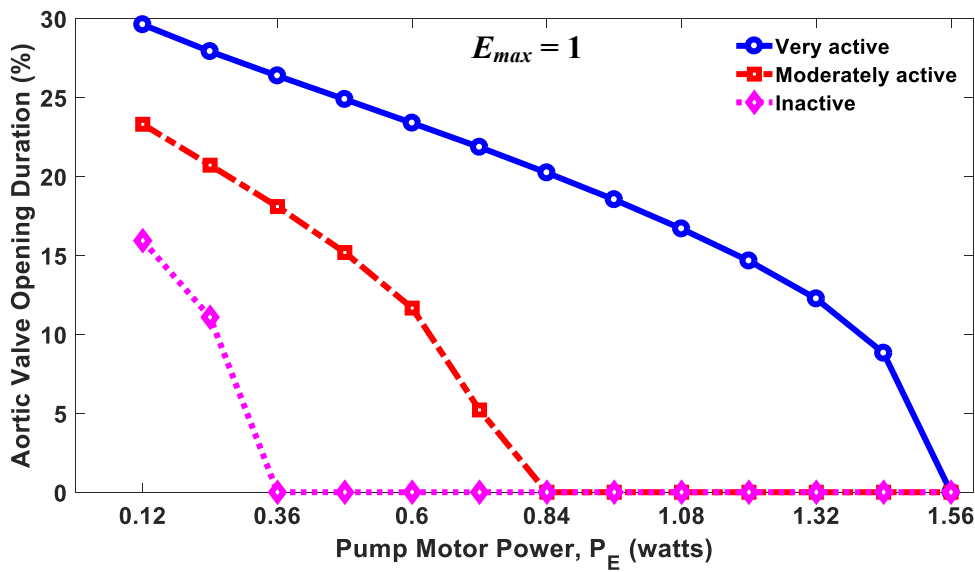

Figure 10: Aortic valve opening duration with respect to the level of physical activity and pump power for a mild heart failure patient ($E_{max}$ = 1)

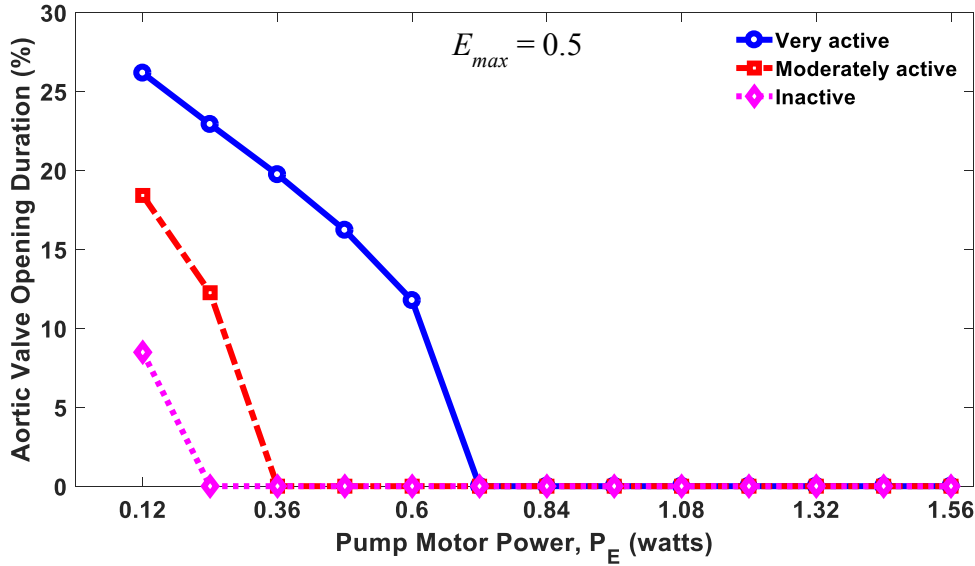

Figure 11: Aortic valve opening duration with respect to the level of physical activity and pump power for a severe heart failure patient ($E_{max} = 0.5$)

As seen in Figures 10 and 11, aortic valve opening duration highly depends on the level of activity for a patient and the severity level of heart failure. For example, as seen in Figure 10, when the patient is very active, the aortic valve can remain open in a cardiac cycle over a wide range of pump motor power. In contrast, the aortic valve can be easily taken over by the pump when a patient is in inactive state and when the pump power is higher than 0.36 *W*. In addition, it was found that the operating range of the pump is much smaller, when the level of heart failure is severe. For example, as seen in Figure 11, for a very active patient with a severe HF, the aortic valve will be closed when the pump power is larger than about 0.7 *W*. However, the aortic valve can remain open for the whole range of the pump power for a very active mild HF patient (see Figure 10). Thus, the *breakpoint*, where the aortic valve closes permanently, can be determined with respect to the level of physical activity as well as the severity of HF. These observations can provide useful information when the pump motor power is used as a control variable. For example, it is difficult to control the pump power to maintain the aortic valve open for a severe heart failure patient, since the allowable operating range of the pump power to keep the aortic valve open is small. This may reduce the cardiac perfusion required to meet the physiological demands. For severe HF patients, an appropriate controller can be developed by considering the severity of HF and by taking into account possible levels of activity, which finds a trade-off between the desired cardiac output and the aortic valve opening duration.

To validate the results, the aortic valve opening duration was also compared with previously reported work. As mentioned in [35], the average opening duration of the aortic valve is 30.5% for a native healthy heart without LVAD, 27% for a mild HF patient with LVAD, and 25% for a severe HF patient with LVAD. This clearly shows that our results are in good agreement with these clinical observations. In addition, it was found that in this work the aortic valve opening duration is highly related to the level of activity and the severity of the native heart.

In addition, Tables in *Appendix B* show the confidence interval of the aortic valve opening duration. It was found that the variance in the aortic valve opening duration increases, as the mean value of $R_s$ decrease and as the pump power increases. For example, as seen in Table C-1, for an active and mild heart failure patient (i.e., $R_s = 0.5$ mmHg/ml/s), the variations in the aortic valve opening duration increase as the pump power increases. It was found that the variation in aortic valve opening duration is about 2 percent point of the mean values on average for these possible pump powers shown in Table C-1. However, for an active and severe heart failure patient as seen Table C-2 ($R_s$ = 0.5 mmHg/ml/s), the average of the variation in the aortic valve opening duration is approximately 0.8%.

3.6 Cardiac Output

Cardiac output (CO) is conventionally calculated as the product of the stroke volume (V) and the HR as: $CO = V \times HR$. However, for HF patients with implanted LVADs, the calculation of cardiac output needs to consider the blood ejected by both the native heart and the LVAD pump. To accomplish this, the cardiac output can be calculated by integrating the aortic flow ($x_5$) [8]. For a cardiac cycle, Equations 30~32 can be used to estimate the cardiac output as:

$$V_T = V_P + V_h \tag{30}$$

$$\int_t^{t+t_c} x_5(\xi) = \int_t^{t+t_c} x_6(\xi) + V_h \tag{31}$$

$$CO_T = CO_P + CO_h \tag{32}$$

where $V_T$ in Equation 30 is the total blood volume pumped into aorta, and $V_P$ and $V_h$ are the blood volume ejected by the LVAD pump and the native heart, respectively. The blood volume can be calculated with Equation 31 for a cardiac cycle $t_c$, from which the total cardiac output of the aorta can be approximated with the summation of two cardiac outputs generated by the pump $CO_P$ and the native heart $CO_h$, by multiplying both sides of Equation 31 with the HR. Note that, since the perturbations in the SVR ($R_s$) are considered in this work, both the pump flow and the aortic flow are functions of random event $\xi$ that can be used to approximate the variation in the SVR. Thus, the integration in Equation 31 is calculated over the domain defined by $\xi$. To solve Equation 31, the trapezoidal rule is used in this work. Using the gPC model of state variable such as $x_5$ and $x_6$, it is possible to provide a measure of confidence interval in the cardiac output prediction.

Note that $V_h$ in Equation 31 is the blood volume of the native heart, which can be estimated from a PV-loop using the gPC models. Figure 12 shows the PV-loops of a mild HF patient with different levels of activity. In this case, the pump power is set to 0.12 $W$ and the HR is 75 bpm. As seen in Figure 12, any perturbations in $R_s$ can affect the stroke work, thus affecting the cardiac output.

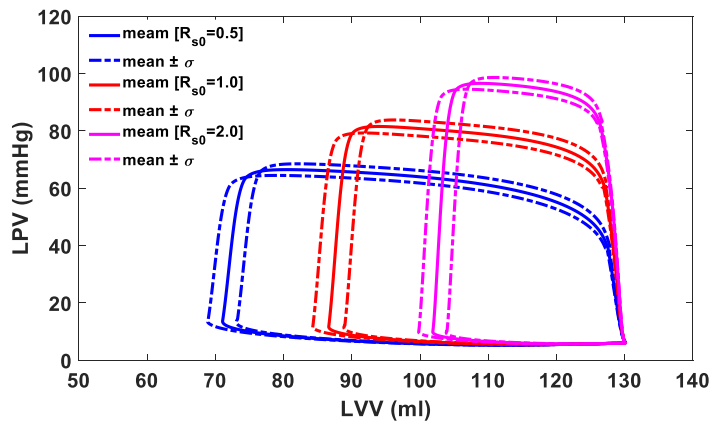

Figure 12: PV-loops generated by the developed stochastic model using gPC theory ($E_{max}$ = 1.0, $P_E$ = 0.12 $W$ and HR = 75 bpm)

In addition, the results of the total cardiac output with respect to different mean values of $R_s$ and pump motor powers are shown in Figures 13 and 14 for a mild HF patient and a severe HF patient, respectively. For simplicity, only three pump powers were investigated, i.e., 0.12, 0.84, and 1.56 *W*, respectively. These values were chosen according to the dynamic behaviour of the aortic valve. For example, the smallest value of the pump power, i.e., 0.12 *W*, is used to ensure the aortic valve can operate normally. The medium value of the pump power, i.e., 0.84 *W*, is used to study the cardiac output when the pump is operated at the *breakpoint* as discussed before. In contrast, the largest value of pump power, i.e., 1.56 *W*, is used to study the effect of perturbations on cardiac output, when the pump takes over the heart function.

In Figures 13 and 14, the vertical *bar* represents the mean value of the cardiac output, while the *error-plot* in each vertical bar represents the variation around the mean value. As seen, the cardiac output and the variation in the cardiac output decrease as the mean value of $R_s$ increases, since the HF patient is less active, and less blood is required. In addition, it was found that the cardiac output increases as the pump power increases. Further, it was found that the variation in the cardiac output decreases as heart function get weaker. For example, as seen in Figure 13, the variation in the cardiac output for an active and mild heart failure patient is about 6.1 L/min, when the pump power is set to 0.12 *W*. In contrast, the variation in cardiac output for an active and severe heart failure is approximately 4.2 L/min with the same pump power as seen in Figure 14.

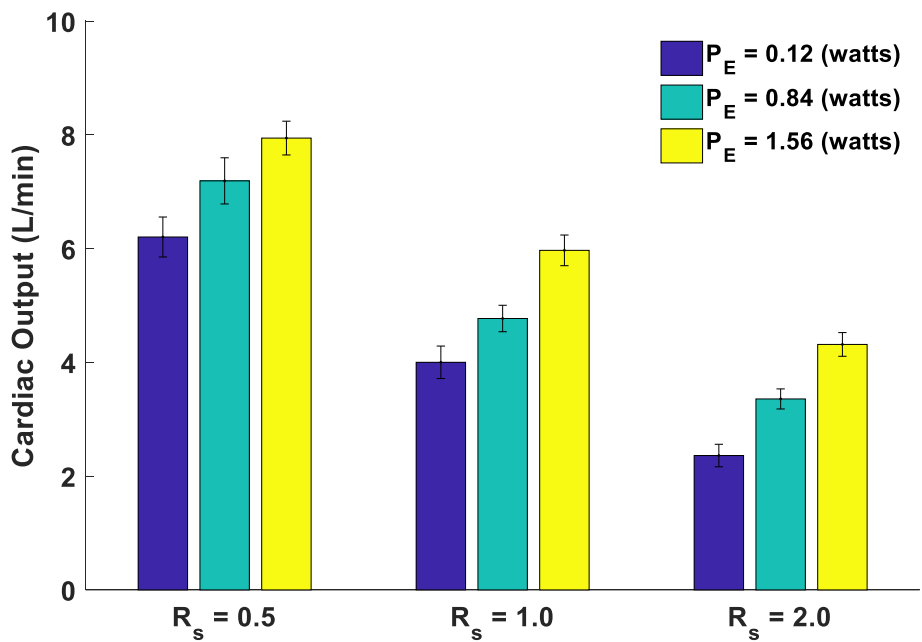

Figure 13: Simulated cardiac output with respect to mean values of $R_s$ and pump motor power for mild heart failure ($E_{max} = 1.0$)

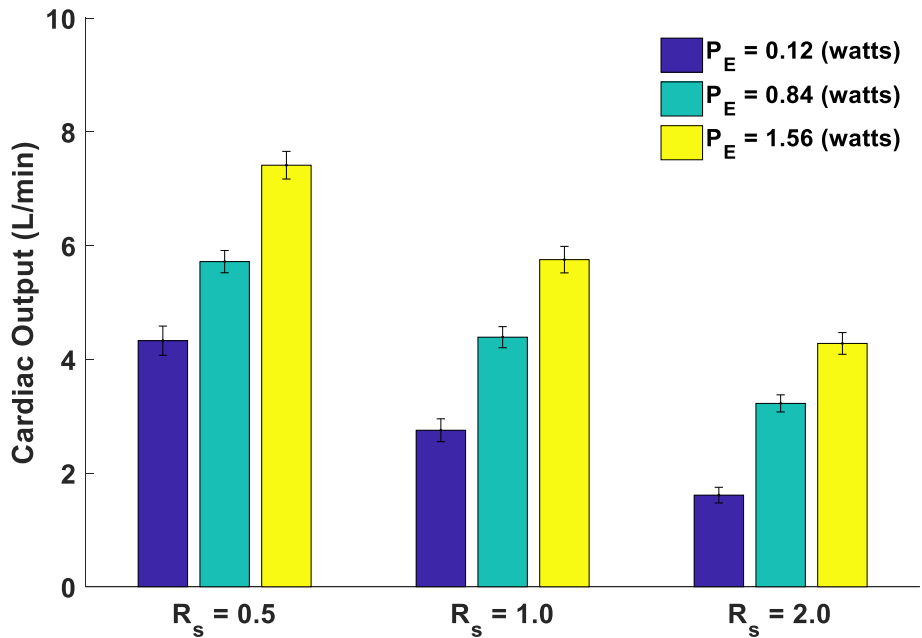

Figure 14: Simulated cardiac output with respect to mean values of $R_s$ and pump motor power for severe heart failure ($E_{max} = 0.5$)

## 4. Conclusions

In this paper, a stochastic cardiovascular-LVAD model is developed to predict the dynamic behaviours of the aortic valve and cardiac output in the presence of uncertainty in the systemic ventricular resistance (SVR). First, the effect of uncertainty in different model parameters on the aortic valve opening duration and the cardiac outputs was evaluated through a half-normal based sensitivity analysis. For the most sensitive model parameter, i.e., the SVR representing the level of activity of a heart failure patient, a generalized polynomial chaos (gPC) expansion was used to approximate the perturbations around a set of mean values of SVR. Further, the effect of uncertainty on the cardiac hemodynamic such as the left ventricular pressure was evaluated using a Galerkin projection. It was found that perturbations in the level of activity (SVR) can significantly affect the aortic valve flow and the pump flow, which can consequently affect the aortic valve opening duration and the total cardiac output in each cardiac cycle. To ensure the proper operation and avoid the permeant closure of the aortic valve, an upper limit of the pump power (i.e., *breakpoint*) is defined. As discussed in the results section, the *breakpoint* of the pump power should be specified with respect to the different levels of activity and different severity of heart failure patients. The more severe the heart failure, the lower value of the pump power. In addition, the variation in the aortic valve opening duration and the cardiac output can be quickly calculated with the gPC model. This is useful for the control design to automatically adjust of the pump power to meet the different physiological demands of human body, since larger variation in the aortic valve opening and cardiac output may weaken the myocardium, which is detrimental to cardiac recovery and can be fatal to HF patients with LVADs. The understanding of the contribution of pump in the overall task of the ejected blood can be very useful for developing a reliable and adaptive controller for the LVAD pump.

## Acknowledgments

Research supported by National Science Foundation (CMMI-1646664, CMMI-1728338, and CMMI-1727487).

## Data availability statement

The stochastic models and the model parameters used to support the finding in this study are included within the article, i.e., Tables 1, and 2, as well as Appendix B.

## *Appendix A*

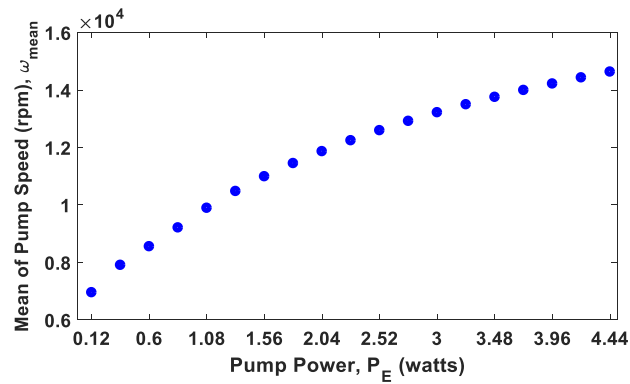

Figure 15 Pump speed corresponding to different pump power levels

## *Appendix B*

The stochastic cardiovascular-LVAD model in the presence of uncertainty in $R_s$.

$$\frac{dx_{10}}{dt} = -\frac{\dot{C}(t)}{C(t)}x_{10} - \frac{1}{C(t)}x_{60} + \frac{1}{C(t)R_M}r(x_{20} \geq x_{10}|(x_{20} - x_{10})) - \frac{1}{C(t)R_A}r(x_{10} \geq x_{40}|(x_{10} - x_{40}))$$

$$\frac{dx_{20}}{dt} = -\frac{1}{C_R R_{S0}}((x_{20} - x_{30})A + (x_{21} - x_{31})B) - \frac{1}{C_R R_M}r(x_{20} \geq x_{10}|x_{20} - x_{10})$$

$$\frac{dx_{30}}{dt} = \frac{1}{C_S R_{s0}}((x_{20} - x_{30})A + (x_{21} - x_{31})B + \frac{1}{C_S}x_{50}$$

$$\frac{dx_{40}}{dt} = -\frac{1}{C_A}(x_{50} - x_{60}) + \frac{1}{C_A R_A}r(x_{10} \geq x_{40}|x_{10} - x_{40})$$

$$\frac{dx_{50}}{dt} = -\frac{1}{L_S}(x_{30} - x_{40} + R_C x_{50})$$

$$\frac{dx_{60}}{dt} = \frac{1}{L^*}(x_{10} - x_{40} - R^* x_{60} + \delta P_E D(t))$$

$$\frac{dx_{11}}{dt} = -\frac{\dot{C}(t)}{C(t)}x_{11} - \frac{1}{C(t)}x_{61} + \frac{1}{C(t)R_M}r(x_{20} \geq x_{10}|(x_{21} - x_{11})) - \frac{1}{C(t)R_A}r(x_{10} \geq x_{40}|(x_{11} - x_{41}))$$

$$\frac{dx_{21}}{dt} = -\frac{1}{C_R R_{s0}}((x_{20} - x_{30})B + (x_{21} - x_{31})C - \frac{1}{C_R R_M}r(x_{20} \geq x_{10}|x_{21} - x_{11})$$

$$\frac{dx_{31}}{dt} = \frac{1}{C_S R_{s0}}((x_{20} - x_{30})B + (x_{21} - x_{31})C + \frac{1}{C_S}x_{51}$$

$$\frac{dx_{41}}{dt} = -\frac{1}{C_A}(x_{51} - x_{61}) + \frac{1}{C_A R_A}r(x_{10} \geq x_{40}|x_{11} - x_{41})$$

$$\frac{dx_{51}}{dt} = -\frac{1}{L_S}(x_{31} - x_{41} + R_C x_{51})$$

$$\frac{dx_{61}}{dt} = \frac{1}{L^*}(x_{11} - x_{41} - R^* x_{61} + \delta P_E E(t))$$

where $A$, $B$, $C$, $D$, and $E$ are the gPC model parameters, $D(t)$ and $E(t)$ are time-varying constants at each time interval. All parameters are calculated with the Galerkin Projection as explained in section 2.2. In these equations above, $x_{i0}$ represents the mean value of the $i^{th}$ physiological variables defined in Table 2, and $x_{i1}$ is the higher order gPC coefficient used to approximate uncertainty due to variations in $R_s$.

## Appendix C

Summary of the variations in the aortic valve opening duration due to perturbations in SVR.

Table C-1: Aortic valve opening duration calculated from gPC coefficients for a mild heart failure patient

| Pump power, $P_E$ (watts) | 0.12 | 0.36 | 0.6 | 0.84 | 1.08 | 1.32 | 1.56 |
| --- | --- | --- | --- | --- | --- | --- | --- |
| $R_s = 0.5$ Opening duration (%) | 29.70±0.80 | 26.47±0.94 | 23.51±1.08 | 20.40±1.23 | 16.88±1.41 | 12.55±1.70 | 0±9.20 |
| $R_s = 1.0$ Opening duration (%) | 23.39±0.74 | 18.23±0.91 | 13.15±1.24 | - | - | - | - |
| $R_s = 2.0$ Opening duration (%) | 16.04±0.66 | 0±4.66 | - | - | - | - | - |

Table C-2: Aortic valve opening duration calculated from gPC coefficients for a severe heart failure patient

| Pump power, $P_E$ (watts) | 0.12 | 0.36 | 0.6 | 0.84 | 1.08 | 1.32 | 1.56 |
| --- | --- | --- | --- | --- | --- | --- | --- |
| $R_s = 0.5$ Opening duration (%) | 26.27±0.85 | 19.89±0.74 | 12.06±1.03 | - | - | - | - |
| $R_s = 1.0$ Opening duration (%) | 18.54±0.57 | 0±1.62 | - | - | - | - | - |
| $R_s = 2.0$ | | | | | | | |

| Opening duration (%) | 8.65±0.57 | - | - | - | - | - |